\documentclass[preprint,showpacs,preprintnumbers,amsmath,amssymb]{revtex4}

\usepackage{graphicx}
\usepackage{dcolumn}
\usepackage{bm}
\usepackage{epsfig}
\usepackage[T1]{fontenc}
\usepackage{ae,aecompl}

\begin{document}

\title{Simulation study of shock reaction on porous material}
\author{Aiguo Xu, Guangcai Zhang, X. F. Pan, and Jianshi Zhu}
\affiliation{National Key Laboratory of Computational Physics, \\
Institute of Applied Physics and Computational Mathematics, P. O.
Box 8009-26, Beijing 100088, P.R.China}
\date{\today}
\begin{abstract}

Direct modeling of porous materials under shock is a complex issue.
We investigate such a system via the newly developed material-point
method. The effects of shock strength and porosity size are the main
concerns. For the same porosity, the effects of mean-void-size are
checked. It is found that, local turbulence mixing and volume
dissipation are two important mechanisms for transformation of
kinetic energy to heat.
 When the porosity is very small, the shocked portion may arrive at a
dynamical steady state; the voids in the downstream portion reflect
back rarefactive waves and result in slight oscillations of mean
density and pressure; for the same value of porosity, a larger
mean-void-size makes a higher mean temperature. When the porosity
becomes large, hydrodynamic quantities vary with time during the
whole shock-loading procedure: after the initial stage, the mean
density and pressure decrease, but the temperature increases with a
higher rate. The distributions of local density, pressure,
temperature and particle-velocity are generally non-Gaussian and
vary with time. The changing rates depend on the porosity value,
mean-void-size and shock strength. The stronger the loaded shock,
the stronger the porosity effects. This work provides a supplement
 to experiments for the very quick procedures and reveals more
fundamental mechanisms in energy and momentum transportation.

\end{abstract}

 \maketitle

\section{Introduction}
Porous materials have extensive applications in industrial and
military fields as well as in our daily life. For example, people
have long been using porous material to shield delicate objects, to
protect things from impact. The porosity characteristics of the
material may significantly influences its dynamical and
thermodynamical behaviors. When a porous material is shocked, the
cavities inside the sample may result in jets and influence its back
velocity\cite{n1}.  Cavity nucleation due to tension waves controls
the spallation behavior of the material\cite{n3}. Cavity collapse
plays a prominent role in the initiation of energetic reactions in
explosives\cite{B2002}.
Most studies on shocked porous materials in literature were experimental\cite%
{P1,Porous1,Porous2,Porous4,Gray2003,Porous8} and theoretical
investigations\cite{Pastine1970,Porous3,Porous5,Porous6,Porous8,WuJing1995,WuJing1996,GengWuTanCaiJing}.
Most of them were focused on the Hugoniots and equations of state.
Due to the inhomogeneities of the material, the underlying
thermodynamical processes in the shocked body are very complex and
far from well-understanding. Understanding these processes plays a
fundamental role in the field and may present helpful information in
material preparation.

From the simulation side, molecular dynamics can discover some
atomistic mechanisms of shock-induced void
collapse\cite{Porous7,Yang}, but the spatial and temporal scales it
may cover are too small compared with experimentally measurable
ones. When treating with the dynamics of structured and/or porous
materials, traditional simulation methods, both the Eulerian and
Lagrangian ones, encountered severe difficulties. The material under
investigation is generally highly distorted during the collapsing of
cavities. The Eulerian description is not convenient to tracking
interfaces. When the Lagrangian formulation is used, the original
element mesh becomes distorted so significantly that the mesh has to
be re-zoned to restore proper shapes of elements. The state fields
of mass density, velocities and stresses must be mapped from the
distorted mesh to the newly generated one. This mapping procedure is
not a straightforward task, and introduces errors. In this study, we
will use a newly developed mixed method, material-point method, to
investigate the shock properties of porous materials.

The material-point method was originally
 introduced in fluid dynamics by Harlow, et
al\cite{H1964} and extended to solid mechanics by Burgess, et al
\cite{MPM}, then developed by various researchers, including
us\cite{JPCM2007,CTP2008,JPD2008}. At each time step, calculations
consist of two parts: a Lagrangian part and a convective one.
Firstly, the computational mesh deforms with the body, and is used
to determine the strain increment, and the stresses in the sequel.
Then, the new position of the computational mesh is chosen
(particularly, it may be the previous one), and the velocity field
is mapped from the particles to the mesh nodes. Nodal velocities are
determined using the equivalence of momentum calculated for the
particles and for the computational grid. The method not only takes
advantages of both the Lagrangian and Eulerian algorithms but makes
it possible to avoid their drawbacks as well.

The following part of the paper is planned as follows.  Section II
presents the theoretical model of the material under consideration.
Section III describes briefly the numerical scheme. Simulation
results are shown and analyzed in section IV. Section V makes the
conclusion.


\section{Theoretical model of the material}

In this study the material is assumed to follow an associative von Mises
plasticity model with linear kinematic and isotropic hardening\cite{CModel}.
Introducing a linear isotropic elastic relation, the volumetric plastic
strain is zero, leading to a deviatoric-volumetric decoupling. So, it is
convenient to split the stress and strain tensors, $\boldsymbol{\sigma }$
and $\boldsymbol{\varepsilon }$, as
\begin{eqnarray}
\boldsymbol{\sigma } &=&\mathbf{s}-P\mathbf{I},P=-\frac{1}{3}\verb|Tr|(%
\boldsymbol{\sigma })\mathtt{,}  \label{PMe1} \\
\boldsymbol{\varepsilon } &=&\mathbf{e}+\frac{1}{3}\theta \mathbf{I},\theta =%
\frac{1}{3}\verb|Tr|(\boldsymbol{\varepsilon })\mathtt{,}  \label{PMe2}
\end{eqnarray}%
where $P$ is the pressure scalar, $\mathbf{s}$ the deviatoric stress tensor,
and $\mathbf{e}$ the deviatoric strain. The strain $\mathbf{e}$ is generally
decomposed as $\mathbf{e}=\mathbf{e}^{e}+\mathbf{e}^{p}$, where $\mathbf{e}%
^{e}$ and $\mathbf{e}^{p}$ are the traceless elastic and plastic components,
respectively. The material shows a linear elastic response until the von
Mises yield criterion,
\begin{equation}
\sqrt{\frac{3}{2}}\left\Vert \mathbf{s}\right\Vert =\sigma _{Y}\mathtt{,}
\label{PM1}
\end{equation}%
is reached, where $\sigma _{Y}$ is the plastic yield stress. The yield $%
\sigma _{Y}$ increases linearly with the second invariant of the plastic
strain tensor $\mathbf{e}^{p}$, i.e.,
\begin{equation}
\sigma _{Y}=\sigma _{Y0}+E_{\tan }\left\Vert \mathbf{e}^{p}\right\Vert
\mathtt{,}  \label{PM4}
\end{equation}%
where $\sigma _{Y0}$ is the initial yield stress and $E_{\tan }$ the
tangential module. The deviatoric stress $\mathbf{s}$ is calculated by
\begin{equation}
\mathbf{s}=\frac{E}{1+\nu }\mathbf{e}^{e}\mathtt{,}  \label{s}
\end{equation}%
where $E$ is the Yang's module and $\nu $ the Poisson's ratio. Denote the
initial material density and sound speed by $\rho _{0}$ and $c_{0}$,
respectively. The shock speed $U_{s}$ and the particle speed $U_{p}$ after
the shock follows a linear relation, $U_{s}=c_{0}+\lambda U_{p}$, where $%
\lambda $ is a characteristic coefficient of material. The pressure $P$ is
calculated by using the Mie-Gr\"{u}neissen state of equation which can be
written as
\begin{equation}
P-P_{H}=\frac{\gamma (V)}{V}[E-E_{H}(V_{H})]  \label{eq-eos}
\end{equation}%
This description consults the Rankine-Hugoniot curve. In Eq.(\ref{eq-eos}), $%
P_{H}$, $V_{H}$ and $E_{H}$ are pressure, specific volume and energy on the
Rankine-Hugoniot curve, respectively. The relation between $P_{H}$ and $%
V_{H} $ can be estimated by experiment and can be written as
\begin{equation}
P_{H}=\left\{
\begin{array}{ll}
\frac{\rho _{0}c_{0}^{2}(1-\frac{V_{H}}{V_{0}})}{(\lambda -1)^{2}(\frac{%
\lambda }{\lambda -1}\times \frac{V_{H}}{V_{0}}-1)^{2}}, & V_{H}\leq V_{0}
\\
\rho _{0}c_{0}^{2}(\frac{V_{H}}{V_{0}}-1), & V_{H}>V_{0}%
\end{array}%
\right.
\end{equation}
In this paper, the transformation of specific internal energy $%
E-E_{H}(V_{H}) $ is taken as the plastic energy. Both the shock compression
and the plastic work cause the increasing of temperature. The increasing of
temperature from shock compression can be calculated as:
\begin{equation}
\frac{\mathrm{d}T_{H}}{\mathrm{d}V_{H}}=\frac{c_{0}^{2}\cdot \lambda
(V_{0}-V_{H})^{2}}{c_{v}\big[(\lambda -1)V_{0}-\lambda V_{H}\big]^{3}}-\frac{%
\gamma (V)}{V_{H}}T_{H}.  \label{eq-eos-temprshock}
\end{equation}%
where $c_{v}$ is the specific heat. Eq.(\ref{eq-eos-temprshock}) can be
resulted with thermal equation and the Mie-Gr\"{u}neissen state of equation%
\cite{explosion}. The increasing of temperature from plastic work can be
calculated as:
\begin{equation}
\mathrm{d}T_{p}=\frac{\mathrm{d}W_{p}}{c_{v}}  \label{eq-eos-temprplastic}
\end{equation}%
Both the Eq.(\ref{eq-eos-temprshock}) and the Eq.(\ref{eq-eos-temprplastic})
can be written as the form of increment.

In this paper the sample
material is aluminum. The corresponding parameters are $\rho_{0}=2700$ kg/m$%
^{3}$, $E=69$ Mpa, $\nu =0.33$, $\sigma _{Y0}=120$ Mpa, $E_{\tan
}=384$ MPa, $c_{0}=5.35$ km/s, $\lambda=1.34$, $c_{v}=880$
J/(Kg$\cdot $K), $k=237$ W/(m$\cdot $K) and $\gamma_0=1.96$ when the
pressure is below $270$ GPa.  The initial temperature of the
material is 300 K.


\section{Outline of the numerical scheme}

As a particle method, the material point method discretizes the
continuum bodies with $N_{p}$ material particles. Each material
particle carries the information of
position $\mathbf{x}_{p}$, velocity $\mathbf{v}_{p}$, mass $m_{p}$, density $%
\rho _{p}$, stress tensor $\boldsymbol{\sigma }_{p}$ , strain tensor $%
\boldsymbol{\varepsilon }_{p}$ and all other internal state
variables necessary for the constitutive model, where $p$ is the
index of particle. At each time step, the mass and velocities of the
material particles are mapped onto the background computational
mesh. The mapped momentum at node $i$ is
obtained by $m_{i}\mathbf{v}_{i}=\sum_{p}m_{p}\mathbf{v}_{p}N_{i}(\mathbf{x}%
_{p})$, where $N_{i}$ is the element shape function and the nodal mass $%
m_{i} $ reads $m_{i}=\sum_{p}m_{p}N_{i}(\mathbf{x}_{p}).$ Suppose
that a computational mesh is constructed of eight-node cells for
three-dimensional
problems, then the shape function is defined as%
\begin{equation}
N_{i}=\frac{1}{8}(1+\xi \xi _{i})(1+\eta \eta _{i})(1+\varsigma
\varsigma _{i})\mathtt{,}  \label{MPMe1}
\end{equation}%
where $\xi $,$\eta $,$\varsigma $ are the natural coordinates of the
material particle in the cell along the x-, y-, and z-directions,
respectively, $\xi _{i}$,$\eta _{i}$,$\varsigma _{i}$ take
corresponding nodal values $\pm 1$. The mass of each particle is
equal and fixed, so the mass conservation equation, $\mathrm{d}\rho
/\mathrm{d}t+\rho \nabla \cdot \mathbf{v}=0$, is automatically
satisfied. The momentum equation reads,
\begin{equation}
\rho \mathrm{d}\mathbf{v/}\mathrm{d}t=\nabla \cdot
\boldsymbol{\sigma }+\rho \mathbf{b}\mathtt{,}  \label{MPMeq2}
\end{equation}%
where $\rho $ is the mass density, $\mathbf{v}$ the velocity, $\boldsymbol{%
\sigma }$ the stress tensor and $\mathbf{b}$ the body force. Equation (\ref%
{MPMeq2}) is solved on a finite element mesh in a lagrangian frame.
Its weak form is
\begin{equation}
\begin{array}{ll}
& \int_{\Omega }{\rho \delta \mathbf{v}\cdot \mathrm{d}\mathbf{v/}\mathrm{d}t%
\mathrm{d}\Omega }+\int_{\Omega }{\delta (\mathbf{v}\nabla )\cdot
\boldsymbol{\sigma }\mathrm{d}\Omega }-\int_{\Gamma _{t}}{\ \delta \mathbf{v}%
\cdot \mathbf{t}\mathrm{d}\Gamma } \\
& -\int_{\Omega }{\ \rho \delta \mathbf{v}\cdot \mathbf{b}\mathrm{d}\Omega }%
=0\mathtt{.}\label{1}%
\end{array}%
\end{equation}%
Since the continuum bodies is described with the use of a finite set
of
material particles, the mass density can be written as $\rho (\mathbf{x}%
)=\sum_{p=1}^{N_{p}}{\ m_{p}\delta (\mathbf{x}-\mathbf{x}_{p})}$, where $%
\delta $ is the Dirac delta function with dimension of the inverse
of volume. The substitution of $\rho (\mathbf{x})$ into the weak
form of the momentum equation converts the integral to the sums of
quantities evaluated at the material particles, namely,
\begin{equation}
m_{i}\mathrm{d}\mathbf{v}_{i}/\mathrm{d}t=(\mathbf{f}_{i})^{\mathrm{int}}+(%
\mathbf{f}_{i})^{\mathrm{ext}}\mathtt{,}  \label{MPMe2}
\end{equation}%
where the internal force vector is given by $\mathbf{f}_{i}{}^{\mathrm{int}%
}=-\sum_{p}^{N_{p}}{m_{p}\boldsymbol{\sigma }}_{p}{\cdot (\nabla
N_{i})/\rho
_{p}}$, and the external force vector reads $\mathbf{f}_{i}{}^{\mathrm{ext}%
}=\sum_{p=1}^{N_{p}}{N_{i}\mathbf{b}_{p}+\mathbf{f}_{i}^{c}}$, where
the vector $\mathbf{f}_{i}^{c}$ is the contacting force between two
bodies. In
present paper, all colliding bodies are composed of the same material, and $%
\mathbf{f}_{i}^{c}$ is treated with in the same way as the internal
force.

The nodal accelerations are calculated by Eq. (\ref{MPMe2}) with an
explicit time integrator. The critical time step satisfying the
stability conditions is the ratio of the smallest cell size to the
wave speed. Once the motion equations are solved on the cell nodes,
the new nodal values of acceleration are used to update the velocity
of the material particles. The strain increment for each material
particle is determined using the gradient of nodal basis function
evaluated at the position of the material particle. The
corresponding stress increment can be found from the constitutive
model. The internal state variables can also be completely updated.
The computational mesh may be the original one or a newly defined
one, choose for convenience,
for the next time step. More details of the algorithm is referred to \cite%
{JPD2008,CTP2008}.

\begin{figure}[tbp]
\centering
\caption{(in JPG format) Snapshots of the shocked porous metal.
$\delta=1.03$, t=250 ns. (a) Contour of pressure, (b) contour of
temperature. The unit of length in this figure is 10 $\mu$m. From
blue to red, the contour value increases. The unit of contour is Mpa
in (a) and is K in (b). The initial velocities of the flyer and
target are $ \pm v_{init} = \pm 1000$ m/s in this case. }
\end{figure}

\begin{figure}[tbp]
\centering
\includegraphics*[scale=0.9,angle=0]{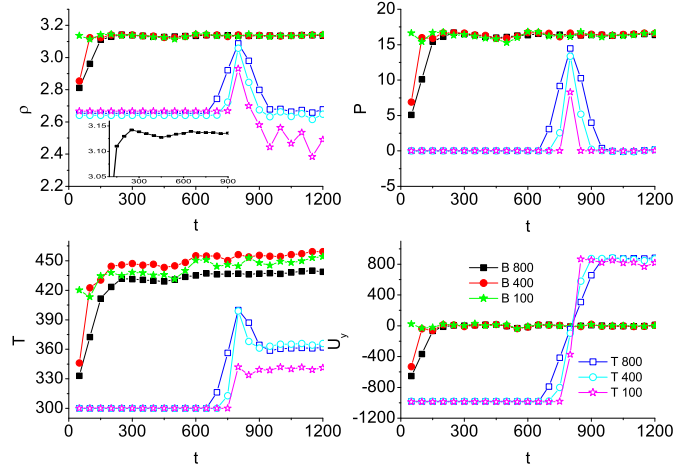}
\caption{(Color online) Variations of mean density, pressure,
temperature and particle velocity with time. The height of the
measured domain are h= 800 $\mu$m, 400$\mu$m and 100 $\mu$m,
respectively, as shown in the legends.  ``B" and ``T" in the legends
means the measured domains are at the bottom and top of the target
body, respectively. The units of density, pressure, temperature,
particle velocity and time are g/cm$^3$, Gpa, K, m/s and ns,
respectively.}
\end{figure}

\begin{figure}[tbp]
\centering
\caption{(in JPG format) Configuration with temperature contour at
time t=1.15 $\mu$s. Other parameters are referred to Fig.1 and
Fig.2. The unit of temperature is K.}
\end{figure}

\section{Results of numerical experiments}

In the present study the porous material is fabricated by a solid
material body with an amount of voids randomly embedded. The
porosity $\delta$ is defined as $\delta = \rho_0 /\rho$, where
$\rho_0$ is the original density of the solid body and $\rho$ is the
mean density of porous material. The porosity $\delta$ in the
simulated system is controlled by the total number $N_{void}$ and
mean size $r_{void}$ of voids embedded. The shock wave to the target
porous metal is loaded via colliding by a second body. For the
convenience of analysis, we set the configurations and velocities of
the two colliding porous bodies symmetric about their impact
interface. The initial velocities of the two colliding bodies are
along the vertical direction and denoted as $\pm v_{init}$. The
impact interface is set at $y = 0$. Periodic boundary conditions are
used in the horizontal directions, which means the investigated real
system is composed of many of the simulated ones aligned
periodically in the horizontal direction. We regarded the upper
porous body as the target, the lower one as the flyer. Compared with
experiments where the target is initially static, the initial
velocity of the flyer is $2 v_{init}$. In this study we focus on the
two-dimensional case. The computational unit is 2 mm in width, as
shown in Fig.1. When we are mainly interested in the loading
procedure of shock wave to porous body, we require that each
simulated body has an enough height so that the rarefactive waves
from the upper and lower free surfaces do not affect the physical
procedure within the time scale under investigation.

 Figure 1 shows two snapshots of
such a process, where Fig.1(a) shows the contour of pressure and
Fig.1(b) shows the contour of temperature. The snapshots show
clearly that, different from the case with perfect solid material,
there is no stable shock wave in the porous materials. When the
compressive waves arrive at a cavity, rarefactive waves are
reflected back and propagate within the compressed portion, which
destroys the original possible equilibrium state there. Even thus,
for the convenience of description, we still refer the compressive
waves to shock waves. Correspondingly, the values of physical
quantities, such as the particle velocity, density, pressure,
temperature, etc, are corresponding mean values calculated in a
region $\Omega$ with $y_1 \leq y \leq y_2$. We will investigate the
effects of initial shock strength and porosity value.

\subsection{Cases with porosity $\delta$=1.03}

We first study the case with $r_{void}$ =50 $\mu$m and the velocity
$v_{init}$ = 1000 m/s, which means the flyer velocity relative to
the target is 2000 m/s. The flyer begins to contact the target at
the time t = 0. Figure 2 shows the variations of mean density,
pressure, temperature and particle velocity with time. These values
are dynamically measured in a bottom and a top domains,
respectively. The height of the target body is 5 mm in this case.
The height of the measured domains are h=800 $\mu$m, 400 $\mu$m and
100 $\mu$m, respectively. For the bottom domain, we choose $y_1 =
100\mu$m. For the top domain, $y_2$ takes the y-coordinate of the
highest material-particle. The lines with solid symbols are measured
values from the bottom and the lines with empty symbols are measured
values from the top. From the figure, we get the following
information: When the shock waves propagate within the bottom domain
$\Omega_b$, the measured mean density, pressure and temperature
increase nearly linearly with time,  up to about t= 150 ns for the
case of h=800 $\mu$m, then further to increase with a decreasing
changing rate. The three quantities arrive at their first maximum
values, 3.14g/cm$^3$, 16.7GPa,  and 432K, at the time t=250 ns. At
this time the shock front has passed the downstream boundary, $y =
810 \mu$m, of the measured domain. (See Fig. 1.) The time delay is
due to dispersion of shock wave in porous media. The followed
concave in either of the $\rho$-,P-,T-curves at about t = 450 ns
shows a downloading phenomenon. The phenomenon is resulted from
rarefactive waves reflected back from the cavities downstream
neighboring to the measured domain. The values of $\rho$ and P
increase and recover to their steady values after that, but the
temperature get a higher value. The secondary loading-phenomenon is
due to the colliding of the upstream and downstream walls during the
collapsing of cavities. Within the following period the density and
pressure keep nearly constants, while the temperature still
increases very slowly. The weak fluctuations in the $\rho$, P, T
curves after $t=650$ ns result from the putting-in of compressive
and rarefactive waves from the two boundaries of the measured domain
$\Omega_b$. The visco-plastic work by these wave series makes the
temperature increase slowly. Since the configurations and velocities
of the flyer and target are symmetric about the plane $y=0$, the
vertical component of particle velocity, $u_y$, is about 0 m/s, the
horizontal component $u_x$ first increases with time, then
oscillates around a small value which is nearly zero. The lines with
empty symbols show that the shock waves arrive at the top free
surface at about t= 800 ns, then rarefactive waves are reflected
back into the target body. Within the time scale shown in the
figure, for the cases with h=800 $\mu$m and 400 $\mu$m, the density
(or pressure) recovers to a value being slightly larger than its
initial one, but the remained temperature is about 60K higher than
the initial temperature and is still increasing; for the case with
h=100 $\mu$m, evident oscillations are found in the curve of density
after t=900 ns. To understand this, we show in Fig.3 the top portion
of the configuration with temperature contour for the time t=1.15
$\mu$s, from which we can find jetting phenomena at the upper free
surface. During the downloading procedure, the top of the porous
body moves upwards with a velocity being about 877 m/s. From the
same data used in Fig.1, we can get the mutual dependence of the
hydrodynamical quantities. The initial transient stage and the final
oscillatory steady state are clearly observable. Due to existence of
the randomly distributed voids, waves with various wave vectors and
frequencies propagate within the shocked sample material. When the
measured domain becomes smaller, more detailed wave structures may
be found. Figure 2 shows clearly this trend.

 It is interesting to check more carefully the procedure of
approaching steady state. Figure 4 shows the standard deviations of
the above four quantities versus time measured in the bottom
domains. It is found that they increase quickly with time at the
very beginning stage, then decrease nearly exponentially to their
steady values. The standard deviation of $u_y$ is larger than that
of $u_x$. The finite sizes of these steady values confirm our
analysis above: what the system arrives is a steady state with local
dynamical oscillations. When the height of the measured domain
increases, the standard deviations of measured quantities become
larger, at least in the transient period.

\begin{figure}[tbp]
\centering
\includegraphics*[scale=0.9,angle=0]{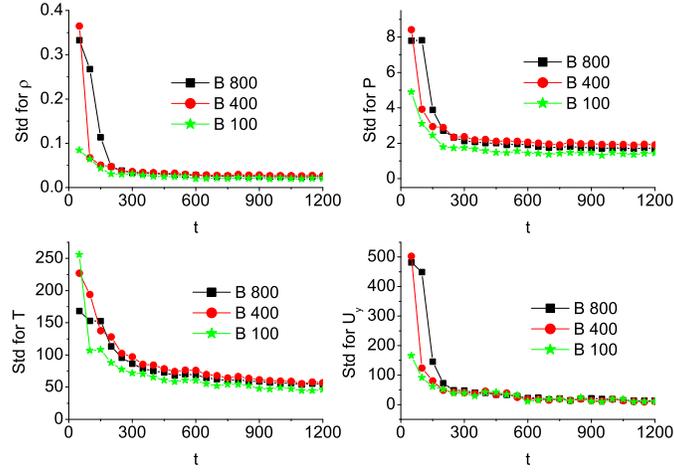}
\caption{(Color online) Standard deviations(Std) of the local
quantities averaged in various spatial scales. The heights of the
measured domains are shown in the legends where ``B" means the
measured domains are at the bottom of the target body. The length
and time units are $\mu$m and ns, respectively.}
\end{figure}

\begin{figure}[tbp]
\centering
\includegraphics*[scale=0.9,angle=0]{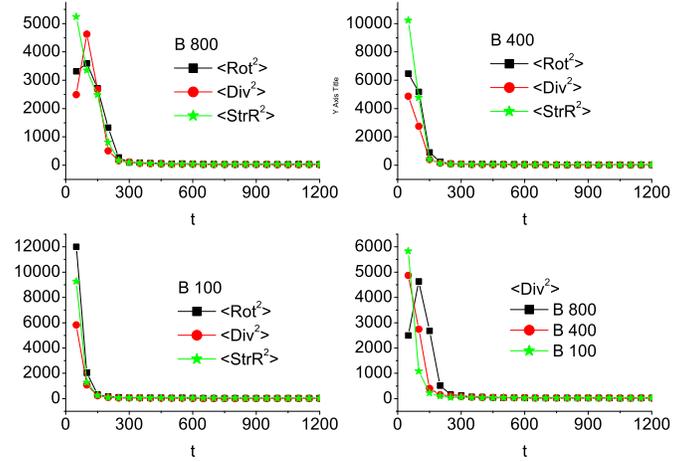}
\caption{(Color online) Variations of the mean values squared of
local rotation, divergence and strain rate with time.  <...> in the
legends denote the mean value of the corresponding quantity and ``B"
means the measured domains are at the bottom of the target body. The
length and time units are $\mu$m and ns, respectively.}
\end{figure}

\begin{figure}[tbp]
\centering
\caption{(in JPG format) Configurations with density contour (a),
pressure contour (b), temperature contour (c) and velocity field (d)
at time t=750 ns. The size of particle velocity is denoted by the
length of arrow timed by 50. The units are the same as in Fig.2.}
\end{figure}

For the case with perfect crystal material, the increase of entropy
result from only from the non-equilibrium procedure of the front of
the shock waves. When cavities exist, the high plastic distortion of
the materials surrounding the collapsed cavities contribute extra
entropy increment. So the local rotation, Rot= $|\nabla \times
\mathbf{u}|$,
 and divergence,
 Div=
 $|\nabla \cdot \mathbf{u}|$,
 make
significance sense in describing shocked porous media. The local
rotation $|\nabla \times \mathbf{u}|$ describes the circular flow
and/or turbulence. The divergence
 $|\nabla \cdot \mathbf{u}|$ describes the changing rate of volume. They show
important mechanisms of entropy and temperature increase in porous
material. The former reflects the turbulence dissipation and the
latter reflects the shock compression. Figure 5 shows the variations
of their mean values squared with time. The behavior of strain rate
$\boldsymbol{\dot{\varepsilon}}$ is plotted as a comparison.
 It is found that all the three quantities decrease
nearly exponentially to their steady state values when shock waves
pass the measured domain $\Omega$. The amplitude of steady strain
rate is very close to that of the rotation. The amplitude of the
divergence is a little larger for this case. Cavity collapse and new
cavitation by the rarefactive waves are the main contributors to the
local divergence. To understand better the fluctuations of the local
density, pressure, temperature, particle velocity and the finite
values of the rotation, divergence, we show in Fig.6 a portion of
the configuration with density contour, pressure contour,
temperature contour and velocity field at time $t=750$ns. In this
case, there is a void around the position (510$\mu$m, 280$\mu$m).

We now checking the effects of the void size. Results for different
void sizes are compared. There is no evident difference in the
steady values of mean density, pressure and particle velocity. But
larger voids contribute to a higher mean temperature. (See Fig.7.)
As for effects on the mean value squared of the local rotation and
divergence, the void size affect only the transient period, but not
the steady values. See Fig.8, where the two cases correspond to
different mean-void-sizes but the same value of porosity,
$\delta=1.03$.
\begin{figure}[tbp]
\centering
\includegraphics*[scale=0.75,angle=0]{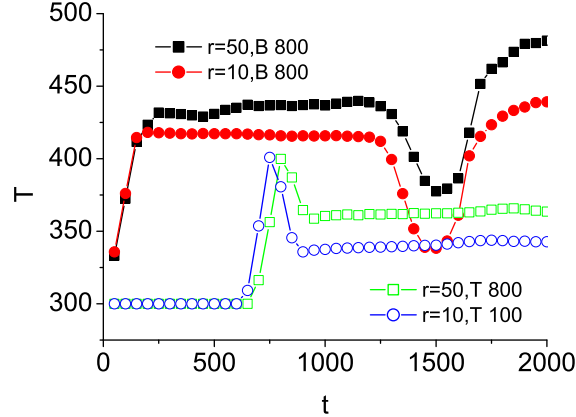}
\caption{(Color online) Effects of the mean void size on the mean
temperature. The mean void size $r$, position and height of the
measured domain are shown in the legend.  ``B" and ``T" means the
measured domains are at the bottom and top of the target body,
respectively. The length and time units are $\mu$m and ns,
respectively.}
\end{figure}

\begin{figure}[tbp]
\centering
\includegraphics*[scale=1,angle=0]{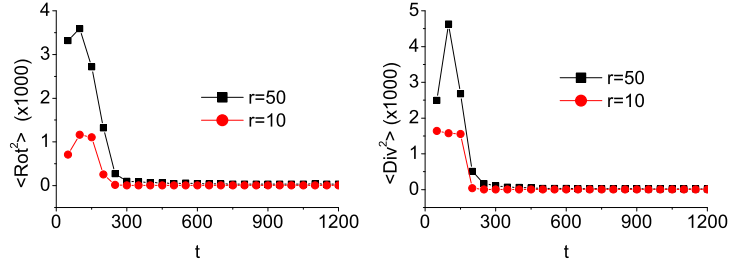}
\caption{(Color online) Effects of mean void size on the mean values
squared of local rotation, divergence. The mean sizes of void are
shown in the legends. The length and time units are $\mu$m and ns,
respectively.}
\end{figure}

\subsection{Cases with porosity $\delta=1.4$}

In this section we study the case with a higher porosity,
$\delta$=1.4. For this case, the mean void size is r=10 $\mu$m.
Figure 9 shows the variations of mean density, pressure, temperature
and particle velocity with time. The initial velocity of the flyer
and the target are $\pm v_{init} = 1000$m/s. The physical quantities
are averaged in a bottom and a top domains. Only the case with h=800
$\mu$m is shown. An evident difference from the low-porosity case
with $\delta=1.03$ is that the mean density and pressure decrease
with time after the initial stage. Correspondingly, the mean
temperature increase with a higher rate. This is due to the
rarefactive waves reflected back from the downstream voids. The
reflected rarefactive waves make the shocked material a little
looser and result in a relatively higher local divergence. The
latter transforms more kinetic energy into heat. At the same time, a
higher porosity means more voids embedded in the material, more
jetting phenomena occur when being shocked. The jetting phenomena
and the hitting of jetted material to the downstream walls of the
voids make a significant increase of local temperature, local
divergence and local rotation. The mean values squared of the local
rotation, divergence and strain rate are shown in Fig.10. These
quantities are measured in the bottom domain with h=800 $\mu$m.
During the initial transient period, the turbulence dissipation
makes the most significant contribution to temperature-increase in
this case. In the later steady state, the three kinds of dissipation
makes nearly the same contribution.

To understand better the inhomogeneity effects in the shocked
portion of the porous body, we show the distributions of density,
pressure, temperature and particle velocity at times t=1200ns,
1250ns and 1300ns in Fig.11. It is clear that these distributions
generally deviate from the Gaussian distribution and vary with time.
In Fig.12 we study the effects of initial impact velocity on the
mean values of the density, pressure and temperature. It is clear
that the decreasing rate of the mean density and the increasing rate
of mean temperature increase when the initial shock wave becomes
stronger. This means that the porosity effects become more
significant when the loaded shock wave becomes stronger.

We now study porosity effects for a fixed shock strength. Figure 13
shows the mean density, and temperature versus time for various
porosities. The initial velocity of the flyer and target are $\pm
v_{init}$ = 1000m/s. When the porosity is very small, the decreasing
rate with time of the mean density becomes higher as the porosity
increases. But when the porosity becomes large, the mean density
show more complex behavior.

\begin{figure}[tbp]
\centering
\includegraphics*[scale=0.9,angle=0]{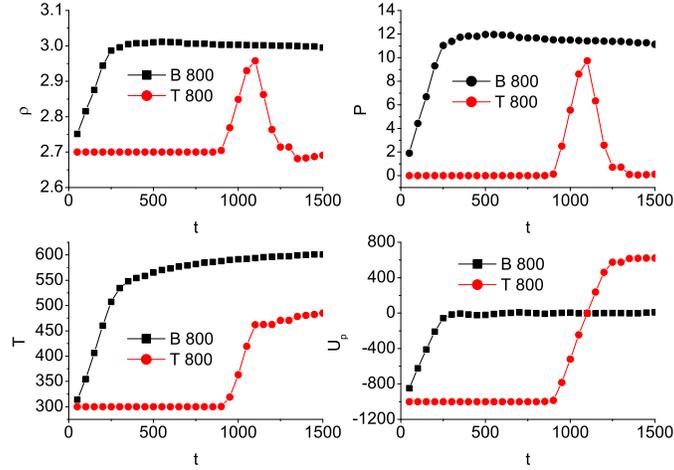}
\caption{(Color online) Variations of mean density, pressure,
temperature and particle velocity with time. Here the porosity
$\delta = 1.4$ and initial flyer velocity relative to the target is
$2 v_{init} = 2000$ m/s. The meanings of ``B", ``T" and units are
the same as in Fig.2.}
\end{figure}
\begin{figure}[tbp]
\centering
\includegraphics*[scale=0.75,angle=0]{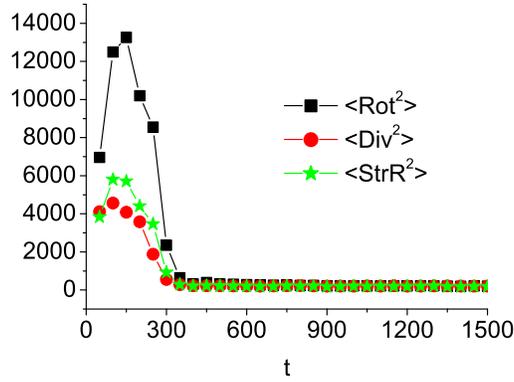}
\caption{(Color online) Variations of the mean values squared of
local rotation, divergence and strain rate with time. The unit of
time is ns. }
\end{figure}

\begin{figure}[tbp]
\centering
\includegraphics*[scale=0.75,angle=0]{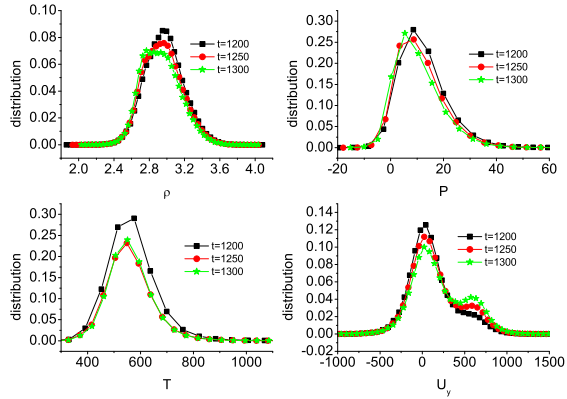}
\caption{(Color online) Distribution of local density, pressure,
temperature, particle velocity at various times. The units are the
same as in Fig.2.}
\end{figure}

\begin{figure}[tbp]
\centering
\includegraphics*[scale=1,angle=0]{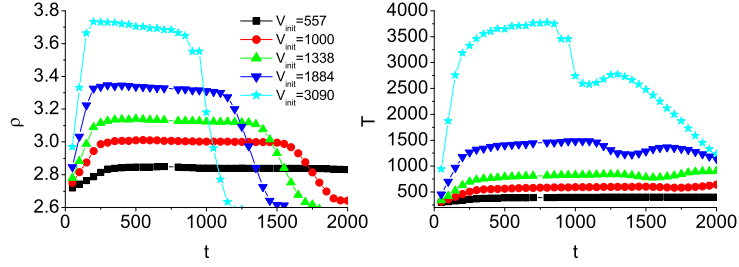}
\caption{(Color online) Mean density and temperature versus time for
various shock strengths. The initial velocity $v_{init}$ are shown
in the legend. The units are the same as in Fig.2. }
\end{figure}

\begin{figure}[tbp]
\centering
\includegraphics*[scale=0.95,angle=0]{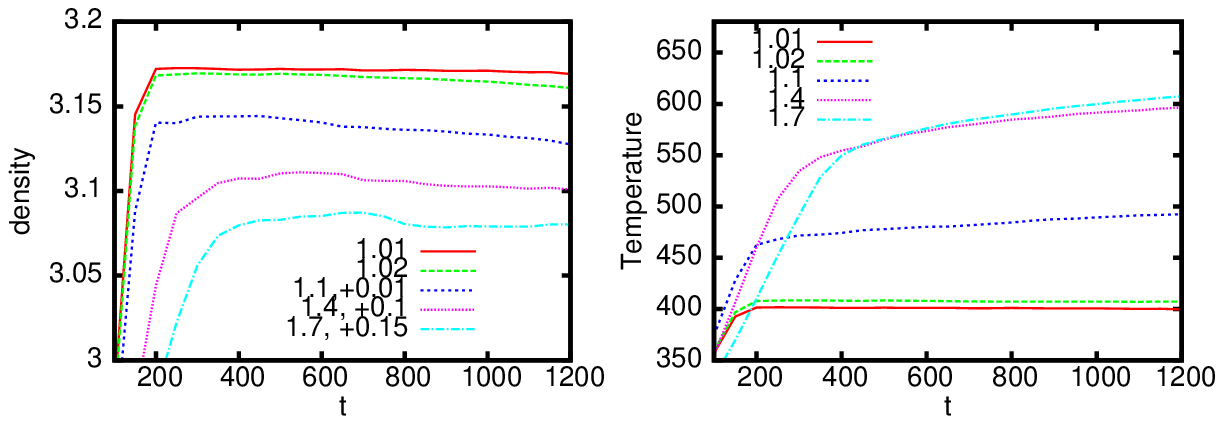}
\caption{(Color online) Mean density and temperature versus time for
various porosities. The values of porosity, 1.01,1.02,1.1,1.4,1.7
are shown in the legends. In the left figure, the lines for cases
with $\delta=$1.1,1.4 and 1.7 are moved upwards by 0.01,0.1, and
0.15, respectively. The units are the same as in Fig.2.}
\end{figure}

\section{Conclusion}

Thermodynamic properties of porous material under shock-reaction is
studied via a direct simulation. The effects of shock strength,
porosity value and the mean-void-size are checked carefully. It is
found that, when the porosity is very small, the shocked portion
will arrive at a dynamic steady state; the voids in the downstream
portion reflect back rarefactive waves and result in slight
oscillations of mean density and pressure; for the same value of
porosity, a larger mean-void-size makes a higher mean temperature.
When the porosity becomes larger, after the initial stage, the mean
density and pressure decrease significantly with time. The
distributions of local density, pressure, temperature and
particle-velocity are generally non-Gaussian and vary with time.
Different from the case with perfect solid material, local
turbulence mixing and volume dissipation exist in the whole loading
procedure and make the system temperature continuously increase. The
changing rates depend on the porosity value, mean-void-size and
shock strength. The stronger the loaded shock, the stronger the
porosity effects. This work is supplementary to experimental
investigations for the very quick procedures and reveals more
fundamental mechanisms in energy and momentum transportation.
\begin{acknowledgments}
We warmly thank Ping Zhang, Jun Chen, Yangjun Ying for helpful
discussions. We acknowledge support by Science Foundations of
Laboratory of Computational Physics, China Academy of Engineering
Physics, and National Science Foundation of China (under Grant Nos.
10702010 and 10775018).
\end{acknowledgments}

\end{document}